# Comparative Study on Millimeter Wave Location-Based Beamforming

Ahmed Abdelreheem, Ahmed M. Nor, Ahmed S. A. Mubarak, Hamada Esmaiel and Ehab Mahmoud Mohamed,
Electrical Engineering Dept., Aswan University, Aswan, Egypt.
{ahmed.abdelreheem, ahmed.nor, ahmed.soliman, h.esmaiel and ehab_mahmoud}@aswu.edu.eg

*Abstract*—this paper presents a comparative study on millimeter wave (mmWave) location-based analog beamforming (BF) techniques based on channel estimation. Localization and compressive sensing (CS) effectively reduces mmWave BF complexity and enhance the performance of mmWave system comparable to the conventional mmWave analog BF techniques. BF techniques based on channel state information (CSI) has high complexity in constructing mmWave channel sensing matrix using CS. Location services based techniques highly reduce this complexity by defining the area within which the user equipment (UE) mostly probable to be exist. In this paper, we study the performance of mmWave location-based BF using various location services. Where, the BF is conducted using channel estimation based CS to estimate both the angle of departures (AoDs) and the angle of arrivals (AoAs) of the mmWave channel.

## I. INTRODUCTION

The wide available spectrum in millimeter wave (mmWave) band makes mmWave a promising solution in the future 5G mobile network to cope with the incredible demand of multi-Gbps data connectivity [1]. However, there are major challenges in mmWave communication channel which highly degrades its performance [2-3], such as poor link budget, high propagation loss and environment effect such as rain and oxygen absorption [3]. Beamforming (BF) technique can be utilized to overcome these issues hence improving the spectral efficiency of mmWave networks. So, exhaustive search BF (EX BF) technique is implemented in IEEE 802.11ad standard as a solution for these challenges. Where, mmWave access point/user equipment (AP/UE) search on all available beam around the UE/AP using the predefined codebook design to select the best pair beams configuration at the transmitter (TX) and the receiver (RX) that maximizing the received power to provide maximum end user data rate [4]. However, this technique suffering from high complexity and low efficiency especially when narrow beams e.g., 5° beamwidth used. Although those sharp beams are recommended to achieve high antenna gain to enhance the system performance, it highly increases the BF complexity. Hence, several techniques have been proposed to overcome EX BF problems, adaptive beamwidth BF techniques, multi-stage codebooks have been presented in [5-8]. Moreover, to enhance the performance of mmWave BF, techniques use the sparse nature of the mmWave channel to estimate the mmWave channel matrix between AP and UE are proposed in [7-10].

Compressive sensing (CS) is one of those techniques [9-11]. Where, the angle of departures (AoDs) and angle of arrivals (AoAs) of the mmWave channel can be estimated using CS. Hence, the best pair beam from the codebook can be selected, and accurately adjusted in the direction of AoDs/AoAs of mmWave channel to increase BF gain. However, CS technique still faces the challenge of constructing the mmWave channel sensing matrix efficiently using a small number of beam switchings to estimate the AoDs and the AoAs of the mmWave channel especially if narrow beams used. Because, selecting wrong configurations of the beams for calculating channel measurements may cause failure in BF due to high errors in the estimated AoDs and AoAs.

Motivated by achieving the main purpose of mmWave BF, selecting a fine mmWave beam pair between AP and UE with a low beam switching complexity which means low overhead in mmWave frame, different localization services based mmWave BF using CS are proposed in literature [10], [11]. where, localization service is used to estimate UE and AP position which used to select best mmWave BF vectors utilized for channel measurements and constructing the mmWave channel sensing matrix since mmWave communication is almost a line of sight (LOS) communication. Then, CS used to predict the AoDs and the AoAs of the channel. Using localization services as assistant to CS in mmWave BF, the complexity of channel state information (CSI) based mmWave BF is highly reduced by guiding the mmWave TX (RX) devices to steer towards the directions that the RX (TX) devices highly probable to be in, respectively. Based on this, the required number of beams to construct the channel sensing matrix is reduced as a result the BF process overall complexity is highly reduced. Also, these schemes guarantee high BF gain.

Without loss of generality, in this paper we compare the performance and the complexity of mmWave location-based BF using CS scheme with various localization accuracy. We focus only on accuracies provided by three wireless communication networks, long term evolution (LTE), Wi-Fi, and Global Position System (GPS) as examples due to a wide spread of these services in wireless communications systems.

The reminder of this paper is organized as follows: Sect. II describes mmWave transmission system. The mmWave BF using CS and location based mmWave BF using CS are presented in Sect. III and Sect. IV, respectively. While, the performance of location based mmWave BF using CS with different location services is compared in Sect. V via numerical simulation using the spectral efficiency and the system complexity as metrics. Finally, Sect. VI concludes this paper.

## II. MMWAVE TRANSMISSION SYSTEM

This study considers mmWave transmission between an AP with a $N_{AP}$ antenna elements and a UE with a $N_{UE}$ antenna elements. Where, the data is up-converted from baseband frequency to radio frequency (RF), then the conventional analog BF is applied to increase the antenna gain. According to the codebook design, the transmitted data are weighted by weight vectors at the TX and RX, as follows [4]:

$$W(n,b) = j^{\text{floor}\left\{\frac{n \times \text{mod}\left(b+\left(\frac{B}{2}\right),B\right)}{\frac{N}{4}}\right\}}, \quad (1)$$
$$n = 0, \ldots, N-1;\ b = 0, \ldots, B-1,$$

where $j = \sqrt{-1}$ and $W(n,b)$ is the antenna weight of antenna element $n$ to achieve BF in the direction of $b$, where $N$ is the total number of mmWave antenna elements and $B$ is the total number of steering beams. It's assumed that both mmWave TX and RX antennas have equal number of steering beams. Also, we considered a geometric channel model with $L$ paths between the AP and UE, which can be written as [9-11]:

$$\mathbf{H} = \frac{1}{\gamma}\sum_{\ell=1}^{L} \mu_\ell \mathbf{p}_{MS}(\varphi_\ell)\mathbf{p}_{BS}^H(\psi_\ell), \quad (2)$$

where $(\ )^H$ is Hermitian (conjugate transpose) and $\mathbf{p}_{AP}(\psi_\ell)$, $\mathbf{p}_{AP}(\varphi_\ell)$ are the array response vectors of the $\ell$ th path at the AP and UE, respectively, which can be written as [11]:

$$\mathbf{p}_{AP}(\psi_\ell) = \left[1, e^{j\frac{2\pi}{\lambda}d\sin(\psi_\ell)}, \ldots, e^{j(N_{AP}-1)\frac{2\pi}{\lambda}d\sin(\psi_\ell)}\right]^T, \quad (3)$$

$$\mathbf{p}_{UE}(\varphi_\ell) = \left[1, e^{j\frac{2\pi}{\lambda}d\sin(\varphi_\ell)}, \ldots, e^{j(N_{UE}-1)\frac{2\pi}{\lambda}d\sin(\varphi_\ell)}\right]^T, \quad (4)$$

where $\gamma$ indicates the average path-loss between AP and UE. $\mu_\ell$, $\varphi_\ell \in [0,2\pi]$ and $\psi_\ell \in [0,2\pi]$ are the complex gain, the azimuth AoD and the azimuth AoA of the $\ell$ th path, respectively. $d$ and $\lambda$ are the distance between the array antenna elements and the mmWave signal wavelength, respectively. Also, 2-D beamforming is considered assuming that all scattering happens in azimuth direction only. The received signal at UE after applying the BF weight vectors in the directions of $b_{tx}$ and $b_{rx}$ at the AP and UE, respectively, can be formulated as:

$$y = \mathbf{W}_{rx}[:,b_{rx}]^H \mathbf{H}\, \mathbf{W}_{tx}[:,b_{tx}]s + \mathbf{W}_{rx}[:,b_{rx}]^H \mathbf{n}, \quad (5)$$

where $y$ and $s$ are the received and transmitted symbols, respectively. $\mathbf{W}_{tx}[:,b_{tx}]$ and $\mathbf{W}_{rx}[:,b_{rx}]$ are the weight vectors of lengths $[N_{AP} \times 1]$ and $[N_{UE} \times 1]$ corresponding to the columns $b_{tx}$ and $b_{rx}$ in the TX cookbook $\mathbf{W}_{tx}$ and the RX codebook $\mathbf{W}_{rx}$, respectively. $\mathbf{H}$ is the mmWave channel matrix of size $N_{UE} \times N_{AP}$ and $\mathbf{n}$ is the $[N_{UE} \times 1]$ Gaussian noise vector corrupting the received symbol. Hence, the optimization problem of the mmWave BF can be formulated as:

$$(b_{tx}^*, b_{rx}^*) = \arg\max_{1 \le b_{tx},b_{rx} \le B} |\mathbf{W}_{rx}[:,b_{rx}]^H \mathbf{H}\, \mathbf{W}_{tx}[:,b_{tx}]|^2, \quad (6)$$

where $b_{tx}^*$ and $b_{rx}^*$ are the TX/RX BF configurations corresponding to the columns of the TX/RX codebooks maximizing the channel BF gain. In the conventional exhaustive searching BF technique, the TX and RX devices search on all available $\mathbf{W}_{tx}[:,b_{tx}]$ and $\mathbf{W}_{rx}[:,b_{rx}]$ pair configurations in the predefined TX and RX codebooks with a beam switching complexity of $B \times B$ as it is assumed that both codebooks have the same total number of steering beams $B$. This technique is independent on estimating the AoDs and AoAs of the channel. Hence, as the number of mmWave TX or RX steering beams increased, i.e., using narrower beamwidth beams in TX or RX, the BF gain of the exhaustive search BF will be improved at the expense of highly increasing the beam switching complexity.

## III. MMWAVE BF USING CS

Motivated by the mmWave channel characteristic especially the sparse nature of channel. MmWave BF can be performed based on CS. Where, CS used to estimate the AoDs and the AOAs of mmWave channel, and as a result building up the channel sensing matrix. Hence, the channel estimation problem can be expressed as a sparse problem [9-11]:

$$\mathbf{Y} = \mathbf{W}_{rx}^H \mathbf{H} \mathbf{W}_{tx}\, \mathbf{S} + \mathbf{W}_{rx}^H \mathbf{n}, \quad (7)$$

where $\mathbf{W}_{tx}$ and $\mathbf{W}_{rx}$ are the TX and the Rx BF matrices with size $[N_{BS} \times B]$ and $[N_{MS} \times B]$, respectively. $\mathbf{S}$ represents a diagonal matrix that carried the transmitted $B$ symbols, here $\mathbf{S} = \sqrt{P}\mathbf{I}_B$, where $P$ is the TX symbol power, and $\mathbf{I}_B$ is the identity matrix with size $B \times B$. While, $\mathbf{W}_{rx}^H \mathbf{n}$ is the noise term. Then transferring the mmWave channel estimation problem to a sparse problem is applied by vectorizing the $\mathbf{Y}$ matrix as:

$$\text{vec}(\mathbf{Y}) = \sqrt{P}\,\text{vec}(\mathbf{W}_{rx}^H \mathbf{H} \mathbf{W}_{tx} + \mathbf{W}_{rx}^H \mathbf{n}), \quad (8)$$

$$\mathbf{y}_v = \sqrt{P}\left(\mathbf{W}_{tx}^T \otimes \mathbf{W}_{rx}^H\right)\text{vec}(\mathbf{H}) + \text{vec}(\mathbf{W}_{rx}^H \mathbf{n}), \quad (9)$$

after that, the channel model given in (2) is used where $\mathbf{y}_v$ can be rewritten as:

$$\mathbf{y}_v = \sqrt{P}\left(\mathbf{W}_{tx}^T \otimes \mathbf{W}_{rx}^H\right)\text{vec}\left(\mathbf{P}_{AP}^* \circ \mathbf{P}_{UE}\right)\boldsymbol{\mu} + \mathbf{n}_x, \quad (10)$$

here $\mathbf{n}_x = \text{vec}(\mathbf{W}_{rx}^H \mathbf{n})$ and $\mathbf{P}_{AP}^* \circ \mathbf{P}_{UE}$ is the Khatri-Rao product of $\mathbf{P}_{AP}^*$ and $\mathbf{P}_{UE}$ resulting in an $N_{BS}N_{MS} \times L$ matrix, where each matrix column $\ell$ represents $\left(\mathbf{p}_{AP}^*(\psi_\ell) \otimes \mathbf{p}_{UE}(\varphi_\ell)\right)$ that is the Kronecker product of AP and UE array responses. To easily estimate the AoDs and the AoAs of the mmWave channel, the AoDs and AoAs values are assumed to be quantized uniformly in a form of grid of $N$ points, where $N \gg L$, as:

$$\overline{\psi_\ell}, \overline{\varphi_\ell} \in \left\{0, \frac{2\pi}{N}, \ldots, \frac{2\pi(N-1)}{N}\right\}, \ell = 1,2,\ldots,L. \quad (11)$$

where $\overline{\varphi_\ell}$ and $\overline{\psi_\ell}$ are the quantized AoDs and AoAs. Of course, this assumption causes a quantization error. Because the AoDs and AoAs are continuous values. Hence, the sparse problem can be formulated as:

$$\mathbf{y}_v = \sqrt{P}\left(\mathbf{W}_{tx}^T \otimes \mathbf{W}_{rx}^H\right)\mathbf{T}_D \mathbf{z} + \mathbf{n}_x, \quad (12)$$

where $\mathbf{z}$ is the vector of length $N^2 \times 1$ that contains the gain of the paths corresponding to the AoDs and AoAs. While, $\mathbf{T}_D$ is a dictionary matrix with size $N_{AP}N_{UE} \times N^2$, where each column in this matrix follows $\left(\mathbf{p}_{AP}(\overline{\psi}_u) \otimes \mathbf{p}_{UE}(\overline{\varphi}_v)\right)$,

$$\overline{\psi}_u = \frac{2\pi u}{N}, u = 0,1,\ldots,N-1, \quad (13)$$

$$\overline{\varphi}_v = \frac{2\pi v}{N}, v = 0,1,\ldots,N-1. \quad (14)$$

The non-zero values of vector $\mathbf{z}$ are the indication of the existence of AoDs and AoAs pairs which mean channel estimation had been successes. Also, it equals to the path gain value of this pair of AoD and AoA. Hence, the vector **z** has only $L$ non-zeros elements and $L \ll N^2$. On the other side, zero values mean there is no channel path in this AoDs and AoAs configuration between TX and RX devices. To formulate the sparse problem completely, an efficient channel sensing matrix $\mathbf{\Phi} = \sqrt{P}\left(\mathbf{W}_{tx}^T \otimes \mathbf{W}_{rx}^H\right)\mathbf{T}_D$ needed to be designed to enable the process of recovering the non-zero elements in **z** vector using small number of beam switching, which is the main issue faced the CS based mmWave BF. To overcome this challenge, BF based on localization using CS is proposed in [10], [11] to enhance the performance of BF gain with low complexity which will be discussed in the following section.

## IV. LOCATION-BASED MmWAVE BF USING CS

User context information is being used widely nowadays as an assistant in wireless communication networks. UE localization is one of this information that can relax mmWave BF process. Localization service is used to provide AP (UE) with UE (AP) estimated position to define the most optimum measurement BF vectors which used to construct an efficient sensing matrix $\mathbf{\Phi}$. This accurate matrix can estimate the AoDs (AoAs) of the channel with low error in angles values. Actually, the AP (UE) estimates a range of quantized AoDs (AoAs) $\bar{\psi}_q$ ($\bar{\varphi}_g$) predicting to be aligned with the actual channel AoDs (AoAs). This range can be expressed as [11]:

$$\bar{\psi}_q = \{\bar{\psi}_u\colon q_1 \leq u \leq q_2, q_1, q_2 \in [0, N-1]\}, \quad (15)$$

$$\bar{\varphi}_g = \{\bar{\varphi}_v\colon g_1 \leq v \leq g_2, g_1, g_2 \in [0, N-1]\}, \quad (16)$$

where $\frac{2\times\pi\times q_1}{N}$ and $\frac{2\times\pi\times q_2}{N}$ are the minimum and maximum values of the estimated AoDs based on the UE location. While, using the estimated location of AP, the minimum and maximum values of the estimated AoAs can be defined as $\frac{2\times\pi\times g_1}{N}$ and $\frac{2\times\pi\times g_2}{N}$. The precoding matrices of TX/RX pairs can be constructed based on $\bar{\psi}_q$ and $\bar{\varphi}_g$ using multiple BF vectors in the range of $[q_1, q_2]$ and $[g_1, g_2]$ for AP and UE, respectively. For AP, assuming $M_{AP}$ different BF vectors with the same beamwidths, each BF vector $m$ in the TX precoding matrices must satisfy:

for $m=1$:

$$\mathbf{W}_{tx}[:,m]\mathbf{p}_{AP}(\bar{\psi}_u) = \begin{cases} C & \text{if } u \in \left[q_1, \frac{q_2-q_1}{M_{AP}}\right], \\ 0 & \text{otherwise} \end{cases}$$

for $2 \leq m \leq M_{AP} - 1$:

$$\mathbf{W}_{tx}[:,m]\mathbf{p}_{AP}(\bar{\psi}_u) = \begin{cases} C & \text{if } u \in \left[m\frac{q_2-q_1}{M_{AP}}, (m+1)\frac{q_2-q_1}{M_{AP}}\right], \\ 0 & \text{otherwise} \end{cases}$$

and for $m=M_{AP}$:

$$\mathbf{W}_{tx}[:,m]\mathbf{p}_{AP}(\bar{\psi}_u) = \begin{cases} C & \text{if } u \in \left[(M_{AP}-1)\frac{q_2-q_1}{M_{AP}}, q_2\right], \\ 0 & \text{otherwise} \end{cases} \quad (17)$$

where $C$ represents a constant value indicating the projection of the BF vector $m$ in the direction of the predefined range of quantized AoDs. Using the same methodology used in above equations, the RX precoding matrix $\mathbf{W}_{rx}$ is constructed by replacing $M_{AP}$, $q_1$ and $q_2$ with $M_{UE}$, $g_1$ and $g_2$, respectively. Then, solving the CS problem. In this study, the orthogonal matching pursuit (OMP) algorithm [9-11] is used to solve such a problem. Hence, the resulted BF vectors corresponding to the estimated AoDs and AoAs are used to transmit mmWave data between AP and UE.

## V. SIMULATION ANALYSIS

In this section, we compare between the three-different location services GPS, LTE, and Wi-Fi, if they used in location based mmWave BF using CS. Where, the localization accuracy of service used in BF is the main player of the efficiency of this technique. A downlink scenario is only considered in this study assuming the mmWave system architecture. Where, the number of AP and UE antenna elements $N_{AP}$ and $N_{UE}$ have the same value equal to 64 using uniform linear array (ULA) antenna.

The channel model used in simulation is Rayleigh fading channel with $L = 3$, and path-loss exponent $n$ equals to 3. While, the AoDs and AoAs assumed to be continuous values in the range of $[0, 2\pi]$ with 28 GHz carrier frequency and 100 MHz bandwidth. Assuming a used localization service provides the AP and the UE with each other locations. Hence, the AP and the UE can estimate the limits of the mmWave channel AoDs and AoAs, $\frac{2\times\pi\times q_1}{N}$, $\frac{2\times\pi\times q_2}{N}$, and $\frac{2\times\pi\times g_1}{N}$, $\frac{2\times\pi\times g_2}{N}$, respectively. Sure, $q_1, q_2$ and $g_1, g_2$ depend on the standard deviation of localization probability and the geometry of the AP and UE estimated positions.

In the simulations, the AP and the UE are randomly existing in the simulation area, and then a random localization error is added to their exact locations for obtaining their estimated positions, as follows:

$$\hat{X}_{BS} = X_{BS} + \delta X_{BS}, \hat{Y}_{BS} = Y_{BS} + \delta Y_{BS}, \quad (18)$$

$$\hat{X}_{MS} = X_{MS} + \delta X_{MS}, \hat{Y}_{MS} = Y_{MS} + \delta Y_{MS}, \quad (19)$$

where $X_{AP}, Y_{AP}$ and $\hat{X}_{AP}, \hat{Y}_{AP}$ are the exact and estimated positions of the AP in the x, y, z coordinates, respectively, and $\delta X_{AP}, \delta Y_{AP}$ are their localization errors. While, $X_{UE}, Y_{UE}, \hat{X}_{UE}, \hat{Y}_{UE}, \delta X_{UE}$ and $\delta Y_{UE}$ are the values for the UE. The values of the localization errors $\delta X_{AP}, \delta Y_{AP}, \delta X_{UE}$ and $\delta Y_{UE}$ are assumed to be uniform random distribution in the range of $[0, 2\sigma]$ m with an average value of $\sigma$ m. Where, $\sigma$ is the standard deviation of localization error in m, which depends on the localization service used in determining the UE or AP estimated position. Based on e of the AP and UE and the maximum expected localization error, the limits $\frac{2\times\pi\times q_1}{N}$, $\frac{2\times\pi\times q_2}{N}$, and $\frac{2\times\pi\times g_1}{N}$, $\frac{2\times\pi\times g_2}{N}$ are adjusted. Also, the beamwidth of the TX/RX BF vectors is adjusted to be $5^o$. Fig.1. shows the spectral efficiency of the system versus SNR which can be defined as [11]:

$$\mathcal{R} = \log_2\left(1 + \frac{P_t|\mathbf{W}_{rx}[:,b_{rx}^*]^H \mathbf{H} \mathbf{W}_{tx}[:,b_{tx}^*]|^2}{\rho^2}\right), \quad (20)$$

where $\rho^2$ is the noise power. This indicates the performance of the system if different wireless netowrks GPS, LTE or Wi-Fi are used to provide the estimated UE and AP positions used in mmWave BF process.

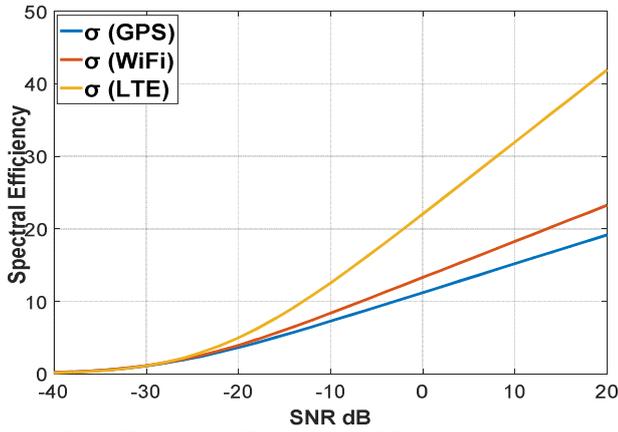
Fig. 1. The spectral efficiency using different location services.

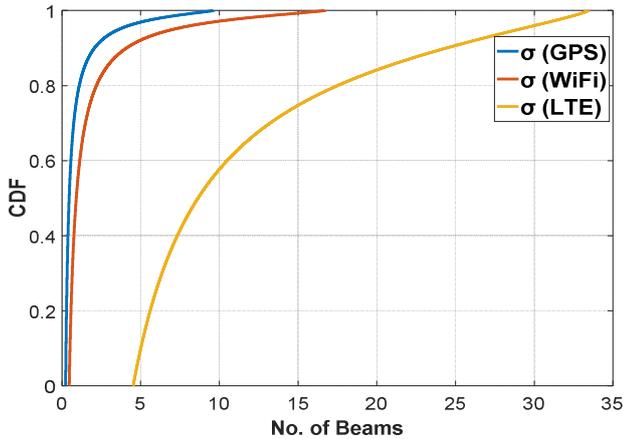
Fig. 2. CDF of number of used TX beams using different location services.

While Fig.2. presents the cumulative distribution function (CDF) of the number of used TX beams. It presents the complexity of location based mmWave BF using CS technique when different location services used. Which shows that to guarantee the optimum performance form GPS, Wi-Fi and LTE, 10, 16 and 33 TX beams is required, respectively.

From Fig.1. and Fig.2., we can clarify that GPS has the lowest number of used TX beams to construct the sensing matrix at the expense of spectral efficiency of the mmWave system which is the lowest one however GPS is not applicable to be used in urban area or NLOS scenarios because of high attenuation faced satellites signals. Also, Wi-Fi guarantees a low complexity BF which is almost same as GPS based technique complexity. In addition, Wi-Fi is more suitable in urban area and NLOS scenarios than GPS. But the spectral efficiency obtained using Wi-Fi is still low nearly half of LTE spectral efficiency. LTE is the more efficient in spectral usage term than GPS and Wi-Fi especially when SNR is larger than -25dB. But with the highest complexity to construct the channel sensing matrices i.e., the TX beams needed by LTE are more than twice the beams required in case of using Wi-Fi or GPS. Because low localization accuracy provided by LTE.

## VI. CONCLUSION

This paper presents a comparative study on location based mmWave BF using CS techniques. where, the AoDs/AoAs searching ranges will be reduced and narrow beams can be used. Hence, the number of BF vectors needed to accurately estimate the actual AoDs/AoAs of the channel is highly reduced. From results, there are a trade-off between the complexity and the spectral efficiency, so when using several location services, some of them can including almost paths of the channel which enhance the spectral efficiency of the system. As when using LTE as a localization service enables the system to achieve the best spectral efficiency nearly twice as spectral efficiency achieved when using GPS or Wi-Fi. On the other side, using GPS or Wi-Fi highly reduces the total system complexity nearly half complexity of LTE based technique at the expense of losing spectrum.

ACKNOWLEDGMENT

This work is partially supported by National Telecom Regulatory Authority (NTRA) Egypt under project title "LTE/WiFi/WiGig internetworking for future 5G cellular networks".

REFERENCES

[1] T. S. Rappaport, S. Sun, R. Mayzus, H. Zhao, Y. Azar, K. Wang, G. N. Wong, J. K. Schulz, M. Samimi, and F. Gutierrez, "Millimeter wave mobile communications for 5G cellular: It will work!," *IEEE Access*, vol. 1, pp. 335–349, 2013.

[2] S. Rangan, S. Member, T. S. Rappaport, and E. Erkip, "Millimeter Wave Cellular Wireless Networks : Potentials and Challenges," vol. 102, no. 3, pp. 1–17, 2015.

[3] W. Roh, J. Y. Seol, J. Park, B. Lee, J. Lee, Y. Kim, J. Cho, K. Cheun, and F. Aryanfar, "Millimeter-wave beamforming as an enabling technology for 5G cellular communications: Theoretical feasibility and prototype results," *IEEE Commun. Mag.*, vol. 52, no. 2, pp. 106–113, 2014.

[4] IEEE Standards Association, IEEE 802.15.3c Part 15.3: Wireless medium access control (MAC) and physical layer (PHY) specifications for high rate wireless personal area networks (WPANs) Amendment 2: Millimeter-wave-based Alternative Physical Layer Extension. 2009.

[5] S. Hur, T. Kim, D. J. Love, J. V Krogmeier, T. a Thomas, and a Ghosh, "Multilevel millimeter wave beamforming for wireless backhaul," *GLOBECOM Work. (GC Wkshps), 2011 IEEE*, pp. 253–257, 2011.

[6] L. Chen, Y. Yang, X. Chen, and W. Wang, "Multi-stage beamforming codebook for 60GHz WPAN," *Proc. 2011 6th Int. ICST Conf. Commun. Netw. China, CHINACOM 2011*, pp. 361–365, 2011.

[7] Ahmed M. Nor and E. M. Mohamed, "Millimeter Wave Beamforming Training Based on Li-Fi Localization in Indoor Environment," *GLOBECOM. (accepted), 2017*.

[8] A. S. A. Mubarak, E. M. Mohamed and H. Esmaiel, "Millimeter wave beamforming training, discovery and association using WiFi positioning in outdoor urban environment," *2016 28th International Conference on Microelectronics (ICM)*, 2016, pp. 221-224, Egypt, 2016.

[9] A. Alkhateeb, O. El Ayach, G. Leus, and R. W. Heath Jr., "Channel estimation and hybrid precoding for millimeter wave cellular systems," IEEE J. Sel. Topics Signal Process., vol. 8, no. 5, pp. 831–846, Oct 2014.

[10] A. Abdelreheem, E. M. Mohamed and H. Esmaiel, "Millimeter wave location-based beamforming using compressive sensing," *2016 28th International Conference on Microelectronics (ICM)*, , pp. 213-216. Egypt, 2016.

[11] A. Abdelreheem, E. M. Mohamed and H. Esmaiel, "Location-Based Millimeter Wave Multi-Level Beamforming Using Compressive Sensing," in *IEEE Communications Letters*, vol. 22, no. 1, pp. 185-188, Jan. 2018.